# An injectable dosimeter for small animal irradiations

**Guy Garty[1], Monica C. Pujol[2] and David J. Brenner[2]**

[1] Radiological Research Accelerator Facility, Columbia University, Irvington, NY, USA
[2] Center for Radiological Research, Columbia University, New York, NY, USA

E-mail: gyg2101@cumc.columbia.edu

## Abstract

We have developed a novel in-vivo dosimeter based on glass encapsulated TLD rods. These encapsulated rods can be injected into mice and used for validating doses to an individual mouse in a protracted irradiation scenario where the mouse is free to move in an inhomogenous radiation field. Data from 30 irradiated mice shows a reliable dose reconstruction within 10% of the nominal delivered dose.

## Introduction

Accuracy and precision in both dosimetry and irradiation experiments is crucial in studies involving animal models [1]. While dose verification in radiotherapy has received much attention and can be partially translated to animal studies (in particular NHP irradiations [2]), placing external radiation monitors on smaller animals provides challenges, in particular when animals are free to move in the irradiator durring protracted irradiations.

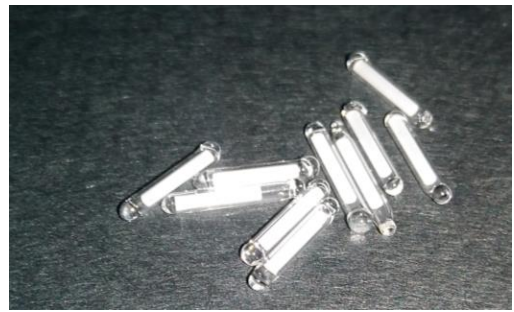

Figure 1: Photo of encapsulated TLD rods.

We have recently developed an ultra low dose rate irradiator for modeling fallout and other environmental contamination exposures. In the VADER (Variable Dose rate Externan irradiatoR) [3] mice are placed in a large cage, placed between movable $^{137}$Cs, sources and are exposed to gamma rays at dose rates of 0.1 to 1.2 Gy/day over a period of several weeks. Clearly in such a case, per-mouse dose validation is critical and restraining mice or providing an external dosimeters (that can be chewed off) are not viable options. This manuscript describes an injectable dosimeter (Fig. 1) based on TLD-100 thermoluminescent dosimeter material, encapsulated in a biocompatible, transparent and heat-stable glass capsule. The encapsiulated TLD rod is slightly smaller than a standard RFID transponder and can thus be injected using standard protocols.

The TLD rod itself is protected from bioliquids within the mouse which may degrade its performance. The encapsulation does not interfere with TLD readout or, importantly, with the annealing process, which requires heating to 400 °C.

We present here the characterisation and calibration of the encapsulated TLD chips and demonstrate their use for *in-vivo* dose verification in mice.

## Materials and Methods

Mouse irradiations and handling were performed under IACUC protocol AC-AAAQ2410.

### Encapsulation of TLD rods

One hundred TLD rods (TLD-100; 1 mm diameter, 6 mm long, from the same batch) were obtained from Thermo Fisher Scientific. Encapsulation was performed by King Precision Glass (Claremont, CA): Briefly, one end of a borosilicate glass tube (KG-33 glass; ID 1.1±0.05 mm; OD 1.5±0.05 mm) was sealed using a propane-oxygen flame. The TLD was inserted and the tube mounted in a heat sink. The free end of the tube was then heated while pulling. After the tube necked off, more heat was applied to finish the seal.

### Irradiation procedures

Prior to each use, the rods were baked and annealed as per the vendor recommendations (Nominally 1 h at 400 °C followed by 2h cooldown to 100°C and 2 h at 100 °C) using an Isotemp programmable muffle furnace (Thermo Fisher Scientific).

For calibration, the rods were placed in an ABS holder and placed on the bottom of a gammacell $^{137}Cs$ irradiator (Canada). Dose rate in the gammacell is tested annually and was between 0.57 and 0.65 Gy/min (depending on sample position) for the measurements reported below.

### Radiation dosimeter chip injection

For *in vivo* irradiations, male 7 week old C57Bl/6J "cage mate" mice were purchased from Charles River Laboratories and kept at the CUIMC animal facility for one week of adaptation. Radiation dosimeter chips were injected at 8 weeks of age and the study performed starting at 9 weeks. Animals were provided with food and water ad libitum and kept on a 12:12 hour light-dark schedule.

Anaesthesia was induced with 2% isoflurane delivered in 100% oxygen for <3 min before the implantation procedure. The encapsulated TLD rods (one per mouse) were placed in a 12 gauge needle coupled with a needle injector (Allflex, USA) and administered by subcutaneous injection in the dorsal neck. Following implantation mice were monitored up to 48 hours for complications.

Mice were irradiated in the same gammacell irradiator, using a standard pie irradiation enclosure.

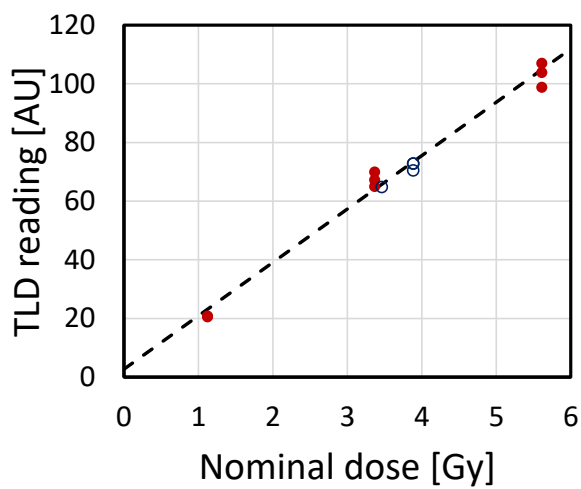

*Figure 2: Sample calibration curve (dashed line) for one of the encapsulated TLD rods. The closed symbols were used to calculate the curve. The Open symbols are subsequent in-vivo measurements using the same rod.*

For mouse irradiations the TLDs were higher in the irradiator (farther from the bottom source), resulting in about 10% lower dose rate, as verified using a NIST traceable Radcal Accudose (10x6-6).

Following irradiation mice were observed for 48h and the TLDs surgically removed.

### Readout

The rods were kept at room temperature for a few days (up to two weeks) after irradiation to allow decay of the low temperature peaks. The rods were read using a Harshaw 2500 TLD reader (Thermo Fisher Scientific), most experiments used a heating profile consisting of a 5 °C/sec ramp up to 300 °C followed by a short hold at 300 °C and cool down to 50 °C.

Collected current on the photomultiplier was integrated starting at a temperature of 180 °C to eliminate the low temperature, time dependant, glow peak.

### Tracking TLD rods

As the response of TLD rods is highly variable, even between rods from the same batch, each TLD rod was assigned a unique code and tracked throught this work. Rods were stored either individually in labled eppendorf tubes or in a custom designed aluminum tray (Protolabs), also used for the anneal process.

## Results

### Calibration curves

Calibration curves were generated individually for each encapsulated TLD rod by performing three consecutive irradiation-read-anneal cycles, at each of the doses 1, 3 and 5 Gy. Independent linear fit parameters were obtained for each rod using linear regression, as implemented by the Microsoft Excel (Redmond, WA) *Slope()* and *Intercept()* functions. A sample fit is shown in Fig. 2. To reconstruct dose, the difference between the intercept and light yield was divided by the slope.

### TLD Tests

In multiple tests of the encapsulated TLDS in both in-vivo and ex-vivo irradiation scenarios resultes were self consistent, with dose reconstructions falling within 5% between detectors and runs. For example, when 47 chips were simultaneously irradiated to 3.9 Gy, the results averaged 3.8Gy with a standard deviation of 0.16 Gy (4%).

Figure 3 shows results of *in-vivo* testing of the TLDs. 3-5 mice per dose were irradiated to doses of up to 4.1 Gy, as described above. The measured doses reproduce the expected ones within 10% in most cases. The standard deviation between TLDs in each grouping was 3-6% of the nominal dose. Dose predictions for unirradiated mice were 0.08±0.06 Gy .

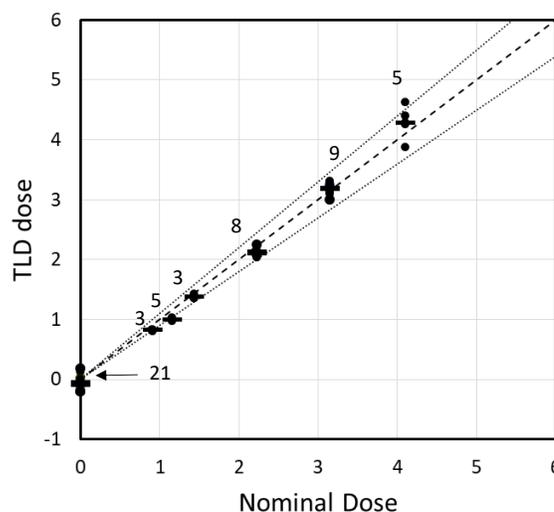

*Figure 3: Dose prediction for 33 measurements and 21 controls. The dashed line represents exact match; the dotted line represents ±10%. The number above each cluster of dots is the number of repeats.*

## Discussion

### *TLD readout*

Even using standard TLD rods, readout requires careful attention to the heating profile. The heating rate is known to influence the observed temperatures of various glow peaks [4] with heating rates above 10°C/min resulting in peaks appearing at higher nominal temperatures.

This temperature lag results from temperature gradients across the heating element, and within the TLD rod and due to improper thermal contact between the heater and TLD [5]. With our encapsulated TLDs, temperature lag effects are expected to be stronger, as the TLD rod is essentially surrounded by an insulator without good thermal contact between the TLD rod and glass or between the glass and heater element.

Corrections for this "temperature lag" have been proposed for heating rates above 10°C/sec (e.g. [6]) and seem to work well [4] but require individual deconvolution of the glow curves. For simplicity we have elected to perform readout at a lower heating rate, minimizing this issue.

We have tested heating profiles, ranging from 1°C/sec to 25°C/sec with and without preheating and did not see significant difference in dose reconstruction between 1 and 10 °C/sec. Therefore, for this work we selected a dose rate of 5°C/sec, balancing the slow heating required to minimize temperature lag and increase reproducibility with the requirement to analyse tens of TLD rods at a time, to support mouse irradiations. At this heating rate we were typically able to read out 10 rods in about an hour.

### *Removing the time dependant glow peak*

Figure 4 shows sample glow curves for encapsulated TLD rods, read out at various times post irradiation. A time dependant glow peak is evident at around 160°C. The light yields at higher temperatures, however independent of time between irradiation and readout.

Figure 5, shows 3 Gy readout for all TLD rods used. It can be seen that while rod to rod variability is on the order of 10%, the variation between consecutive readouts of the same rod (after removing the time dependant glow peak) is typically less than 5%.

### *Sample tracking.*

One of the main problems with TLD chips is that even within a single batch the chip to chip variability is rather poor and in order to perform reliable dosimetry each chip needs to be individually calibrated. This raises the need to keep track of which chip is which.

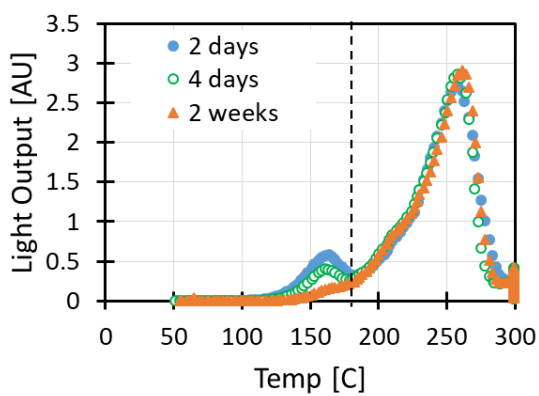

Figure 4: Glow curves for encapsulated TLD-100 irradiated to 3 Gy and read at various times post irradiation. Glow curves were integrated right of the dashed line (180°C).

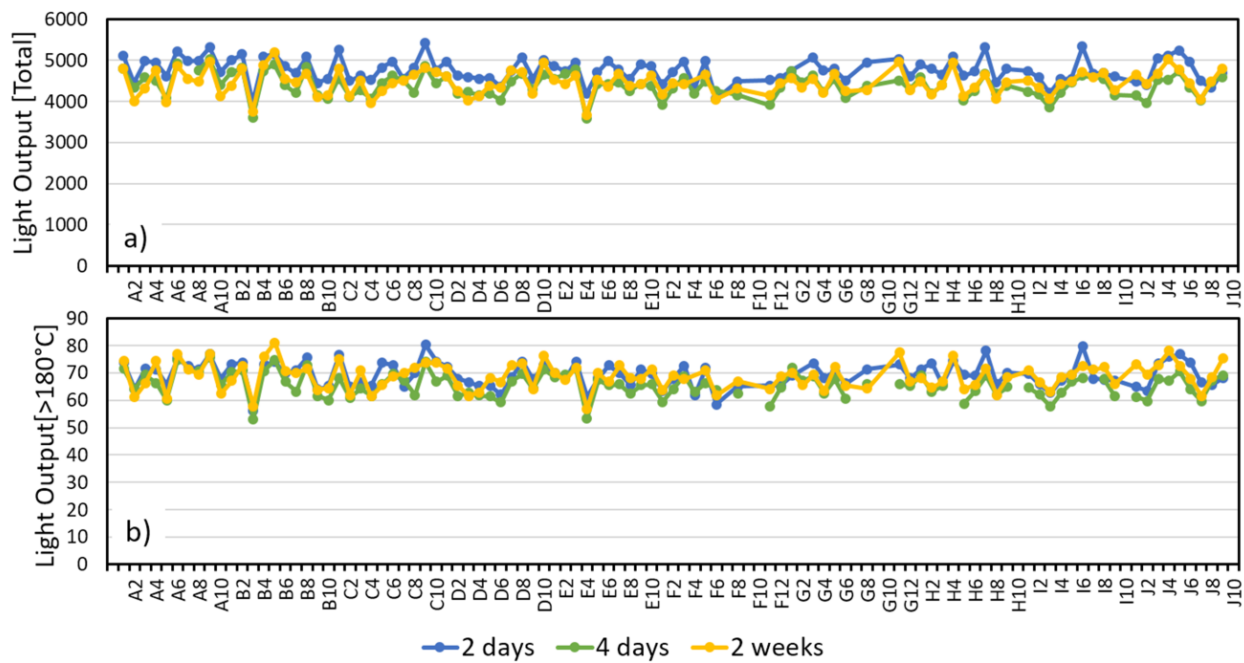

*Figure 5: Light output from all TLDs used irradiated to 3 Gy and read out at various times. a) "Raw" light yield and b) light yield above 180°C.*

In this work we maintained the TLDs in individual labelled tubes although this is far from an ideal solution as it requires constant diligence. We are investigating methods for intrinsically labelling the TLDs but any such method cannot block light and must be able to withstand the 400 °C bake required between irradiations.

## Use for dose reconstruction

Accuracy and precision in both dosimetry and irradiation experiments is crucial in studies involving animal models [7]. In most cases this is achieved by proper characterisation of the irradiator however in cases where the irradiations are long and animals are free to move, this is not practical. The encapsulated TLDs we have developed provide a reliable way to fold the animal movement and validate the total delivered dose.

In tests performed using a Gammacell irradiator, the doses delivered to mice reproduced the nominal delivered dose within better than 10%. Based on the *ex-vivo* results, about half of this variation is due to intrini=sic variability in the TLD rods and dose reconstruction and the other half due to mouse motion within the radiation field.

Ongoing studies are utilizing these detectors for long term (up to 30 day) irradiations of mice in a low dose rate environemnt. Preliminary results indicate that the actual delivered dose matches the target dose with a similar range.

An issue that was detected in these long studies is the loss of an occasional TLD rod. This was resolved with the application of a small drop of Tissue Adhesive to close the wound after the TLD rod injection.

## Conclusions

We presented here a novel encapsulated TLD system for performing in-vivo dosimetry on mice exposed to varying, long term radiation fields. Reconstructed doses were within 10% of the expected dose.

## Acknowledgements

The authors would like to acknowledge the support of Eric King and King Precision Glass Inc. (Claremont, CA) in developing the TLD encapsulation procedures.

This work was supported by grant number U19-AI067773 to the Center for High-Throughput Minimally Invasive Radiation Biodosimetry, from the National Institute of Allergy and Infectious Diseases (NIAID), National Institutes of Health (NIH). The content is solely the responsibility of the authors and does not necessarily represent the official views of the NIAID or NIH.